\documentclass{IEEEtran}

\usepackage{fancyhdr}

\makeatletter
\DeclareOldFontCommand{\rm}{\normalfont\rmfamily}{\mathrm}
\DeclareOldFontCommand{\sf}{\normalfont\sffamily}{\mathsf}
\DeclareOldFontCommand{\tt}{\normalfont\ttfamily}{\mathtt}
\DeclareOldFontCommand{\bf}{\normalfont\bfseries}{\mathbf}
\DeclareOldFontCommand{\it}{\normalfont\itshape}{\mathit}
\DeclareOldFontCommand{\sl}{\normalfont\slshape}{\@nomath\sl}
\DeclareOldFontCommand{\sc}{\normalfont\scshape}{\@nomath\sc}
\makeatother

\usepackage{hyperref}
\usepackage{amsmath,amssymb,amsthm}
\usepackage{algorithm,algorithmic}
\usepackage{booktabs}
\usepackage{graphicx}
\usepackage{color}

\newcommand{\mA}{\mathbf{A}}
\newcommand{\mH}{\mathbf{H}}
\newcommand{\mW}{\mathbf{W}}
\newcommand{\mX}{\mathbf{X}}
\newcommand{\mY}{\mathbf{Y}}
\newcommand{\mS}{\mathbf{S}}

\newcommand{\vy}{\mathbf{y}}
\newcommand{\vx}{\mathbf{x}}
\newcommand{\cY}{\mathsf{Y}}
\newcommand{\cH}{\mathsf{H}}

\newcommand{\cS}{\mathsf{S}}

\newcommand{\vv}{\mathbf{v}}

\newcommand{\rtm}{\textsuperscript{\tiny{\textregistered}}}

\DeclareMathOperator{\argmin}{\mathrm{arg\,min}}
\newcommand{\bydef}{\overset{\textrm{def}}{=}}

\title{Direction of Arrival with One Microphone, a few LEGOs, and Non-Negative Matrix Factorization}
\author{Dalia El Badawy and Ivan Dokmani\'c\thanks{In line with the philosophy of  reproducible research, code and data to reproduce the results of this paper are available at \url{http://github.com/swing-research/scatsense}.},~\IEEEmembership{Member,~IEEE}
\thanks{D. El Badawy is a student at EPFL, Switzerland, e-mail: dalia.elbadawy@epfl.ch.}
\thanks{I. Dokmani\'{c} is with ECE Illinois, e-mail: dokmanic@illinois.edu.}
\thanks{Manuscript received January xx, 2018; revised Month xx, 2018.}}

\begin{document}

\maketitle

\thispagestyle{fancy}

\begin{abstract}
Conventional approaches to sound source localization require at least two microphones. It is known, however, that people with unilateral hearing loss can also localize sounds. Monaural localization is possible thanks to the scattering by the head, though it hinges on learning the spectra of the various sources. We take inspiration from this human ability to propose algorithms for accurate sound source localization using a single microphone embedded in an arbitrary scattering structure. The structure modifies the frequency response of the microphone in a direction-dependent way giving each direction a signature. While knowing those signatures is sufficient to localize sources of white noise, localizing speech is much more challenging: it is an ill-posed inverse problem which we regularize by prior knowledge in the form of learned non-negative dictionaries. We demonstrate a monaural speech localization algorithm based on non-negative matrix factorization that does not depend on sophisticated, designed scatterers. In fact, we show experimental results with ad hoc scatterers made of LEGO bricks. Even with these rudimentary structures we can accurately localize arbitrary speakers; that is, we do not need to learn the dictionary for the particular speaker to be localized. Finally, we discuss multi-source localization and the related limitations of our approach.
\end{abstract}


\section{Introduction}
\label{sec:introduction}
\IEEEPARstart{I}{n} this paper, we present a computational study of the role of scattering in sound source localization. We study a setting in which localization is a priori not possible: that of a single microphone, referred to as monaural localization. It is well established that people with normal hearing localize sounds primarily from binaural cues---those that require both ears. Different directions of arrival (DoA) result in different interaural time differences which are the dominant cues for localization at lower frequencies, as well as in interaural level differences (ILD) which are dominant at higher frequencies \cite{blauert1997}. The latter are linked to the head-related transfer function (HRTF) which encodes how human and animal heads, ears, and torsos scatter incoming sound waves. This scattering results in direction-dependent filtering whereby frequencies are selectively attenuated or boosted; the exact filtering depends on the shape of the head and ears and therefore varies for different people and animals. Thus the same mechanism responsible for frequency-dependent ILDs in the HRTF also provides monaural cues. The question is then, can these monaural cues embedded in the HRTF be used for localization? 

Indeed, monaural cues are known to help localize in elevation \cite{blauert1997} and resolve the front/back confusion \cite{musicant1984}: two cases where binaural cues are not sufficient. Additionally, studies on the HRTFs of cats \cite{cathrtf} and bats \cite{bathrtf} also reveal their use for localization in both azimuth and elevation, albeit in a binaural setting. This implies that the directional selectivity of the HRTF i.e., the monaural cues, is sufficient to enable people with unilateral hearing loss to localize sounds, though with a reduced accuracy compared to the binaural case \cite{oldfield1986}. 

\subsection{Related Work}
Combining HRTF-like directional selectivity with source models has already been explored in the literature  \cite{harris2000,metamaterials,saxena2009,elbadawy2017}. For example, in one study \cite{saxena2009}, a small microphone enclosure was used to localize one source with the help of a Hidden Markov Model (HMM) trained on a variety of sounds including speech. In another study \cite{metamaterials}, a metamaterial-coated device with a diameter of 40 cm and a dictionary of noise prototypes were used to localize known noise sources. In our previous work \cite{elbadawy2017}, we used an omnidirectional sensor surrounded by cubes of different sizes and a dictionary of spectral prototypes to localize speech sources.

A single omnidirectional sensor can also be used to localize sound sources inside a known room \cite{ivanthesis}. Indeed, in place of the head, the scattering structure is then the room itself and the localization cues are provided by the echoes from the walls \cite{roomhelps}. The drawback is that the room should be known with considerable accuracy---it is much more realistic to assume knowing the geometry of a small scatterer.

As for source models, those used in previous work on monaural localization rely on full complex-valued spectra \cite{metamaterials}. Other approaches to multi-sensor localization with sparsity constraints also operate in the complex frequency domain \cite{malioutov2005,boufounos2011,cagli2013}. In this paper, we choose to work with non-negative data which in this case corresponds to the power or magnitude spectra of the audio. We highlight two reasons for this choice. First, unlike the multi-sensor case, the monaural setting generates fewer useful relative phase cues. Second, if \emph{prototypes}---that is, the exact source waveform---are assumed to be known as in \cite{metamaterials}, there are no modeling errors or challenges associated with the phase information. We, however, assume much less, namely only that the source is speech. It is then natural to leverage the large body of work that addresses dictionary learning with real or non-negative values as opposed to complex values. In particular, we consider models based on non-negative matrix factorization (NMF). NMF results in a parts-based representation of an input signal \cite{lee1999} and can for instance identify individual musical notes \cite{fevotte2009}. Thus with training data, NMF can be used to learn a representation for each source \cite{usm,schmidt2006}. For more flexibility, it can also be used to learn an overcomplete dictionary where each source admits a sparse representation \cite{usm,schmidt2006}. For the latter, either multiple representations are concatenated \cite{usm} or the learning is modified by including sparsity penalties \cite{schmidt2006,leroux2015}. 
 
To solve the localization problem, we first fit the postulated non-negative model to the observed measurements. The cost functions previously used often involve the Euclidean distance \cite{metamaterials,elbadawy2017,boufounos2011,malioutov2005,cagli2013}. Non-negative modeling lets us use other measures more suitable for speech and audio such as the Itakura--Saito divergence \cite{fevotte2009}. While NMF is routinely used in single-channel source separation \cite{usm,virtanen2007,smaragdis2007,dikmen2009}, speech enhancement \cite{mohammadiha2013}, polyphonic music transcription \cite{smaragdis2003}, and has been used in a multichannel joint separation and localization scenario \cite{traa2015}, the present work is to the best of our knowledge the first time NMF is used in single-channel source localization. Finally, when the localization problem is ill-posed, as is the case for the monaural setting, various regularizations are utilized. Typical regularizers promote sparsity \cite{metamaterials}, group sparsity \cite{boufounos2011,cagli2013} or a combination thereof \cite{elbadawy2017}.

\subsection{Contributions \& Outline}
\label{sub:objectives}
The current paper extends our previous work \cite{elbadawy2017} in several important ways. We summarize the contributions as follows:
\begin{itemize}
  	\item We derive an NMF formulation for monaural localization via scattering;
    \item We formulate two different regularized cost functions with different distance measures in the data fidelity term to solve the localization based on either universal or speaker-dependent dictionaries;
    \item We present extensive numerical evidence using simple ``devices'' made from LEGO\rtm bricks;
    \item For the sake of reproducibility, we make freely available the code and data used to generate the results.
\end{itemize}
Unlike \cite{saxena2009}, the source model we present easily accommodates more than one source. And unlike \cite{harris2000} or \cite{metamaterials}, we present localization of challenging sources such as speech without the need for metamaterials or accurate source models---we only use ad hoc scatterers and NMF. In this paper we limit ourselves to anechoic conditions and localization in the horizontal plane as our goal is to assess the potential of this simple setup.

In the following, we first lay down an intuitive argument for how monaural cues help as well as a simple algorithm for localizing white sources. We then formulate the localization problem using NMF and give an algorithm for general colored sources in Section \ref{sec:main}. In Section \ref{sec:exp}, we describe our devices and results for localizing white noise and speech. 

\section{Background}
The sensor we consider in this work is a microphone, possibly omnidirectional, embedded in a compact scattering structure; we henceforth refer to it as ``the device''. We discretize the azimuth into $D$ candidate source locations $\Omega = \{ \theta_1, \theta_2, \ldots, \theta_D \}$ and consider the standard mixing model in the time domain for $J$ sources incoming from directions $\Theta = \{ \theta_j \}_{j \in \mathcal{J}}$,
\begin{equation}
y(t) = \sum_{j \in \mathcal{J}}  s_j(t)\ast h_j(t) + e(t),
\label{eq:mixt}
\end{equation}
where $\mathcal{J} \subseteq \{ 1, 2, \ldots, D\} \overset{\textrm{def}}{=} \mathcal{D}$, $|\mathcal{J}| = J$, $\ast$ denotes convolution, $y$ is the observed signal, $s_j$ is the $j^{\textrm{th}}$ source signal, $h_j(t) \overset{\textrm{def}}{=} h(t;\theta_j)$ is the impulse response of the directionally-dependent filter, and $e$ is additive noise. The goal of localization is then to estimate the set of directions $\Theta$ from the observed signal $y$. Note that in general we could also include the elevation by considering a set of $D$ directions in 3D, though this would likely yield many additional ambiguities. 

The mixing \eqref{eq:mixt} can be approximated in the short-time Fourier transform (STFT) domain as
\begin{equation}
Y(n,f)  = \sum_{j \in \mathcal{J}}  S_j(n,f) H_j(f) + E(n,f),
\label{eq:mixstft}
\end{equation}
where $n$ and $f$ denote the time and frequency indices. This so-called narrowband approximation holds when the filter $h_j$ is short enough with respect to the STFT analysis window \cite{kowalski2010,parra2000}. For reference, the impulse response corresponding to an HRTF is around 4.5 ms long \cite{cipic}, while the duration of the STFT window for audio is commonly anywhere between 5 ms and 128 ms during which the signal is assumed stationary. Finally, the mixture's spectrogram with $N$ time frames and $F$ frequency bins can be written as
\begin{equation}
  \cY = \sum_{j \in \mathcal{J}} \textrm{diag}(\cH_j)\cS_j + \mathsf{E},
\label{eq:mixspec}
\end{equation}
where $\cY \in \mathbb{C}^{F\times N }$, $\cS_j \in \mathbb{C}^{F\times N }$ the spectrogram of the source impinging from $\theta_j$, $\cH_j \in \mathbb{C}^{F}$ is the frequency response of the directionally-dependent filter, $\mathsf{E} \in \mathbb{C}^{F\times N }$ is the spectrogram of the additive noise, and $\textrm{diag}(\vv)$ is a matrix with $\vv$ on the diagonal.

\begin{figure}[t]
\begin{minipage}[b]{0.5\linewidth}
  \centering
   \centerline{\includegraphics[width=\linewidth]{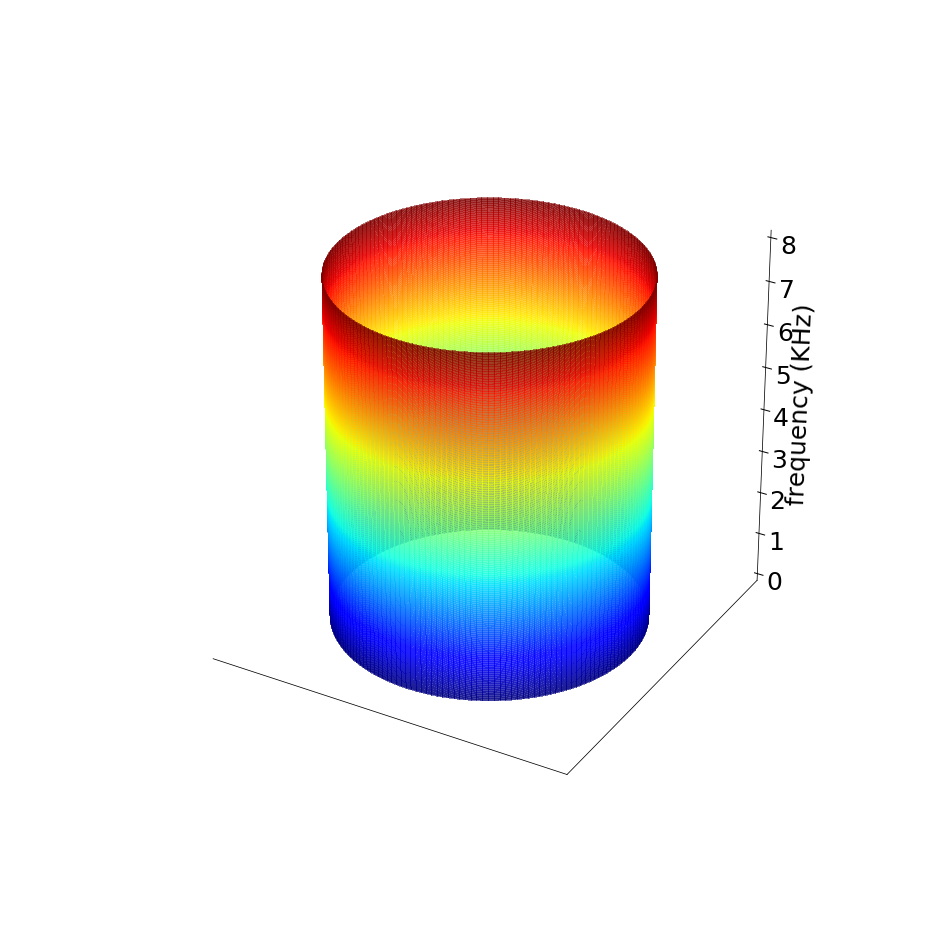}} 
  \centerline{(a) No scattering}
\end{minipage}%
\begin{minipage}[b]{0.5\linewidth}
\centering
   \centerline{\includegraphics[width=\linewidth]{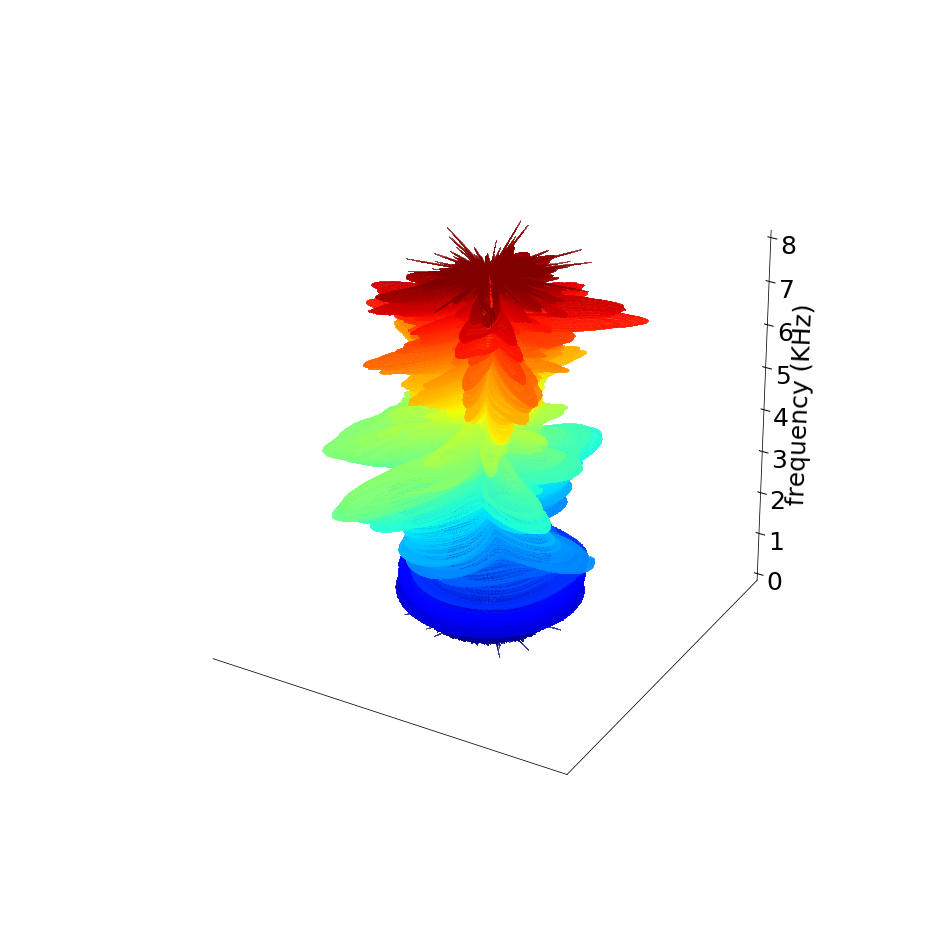}} 
  \centerline{(b) LEGO1}
\end{minipage}
\begin{minipage}[b]{0.5\linewidth}
  \centering
  \centerline{\includegraphics[width=\linewidth]{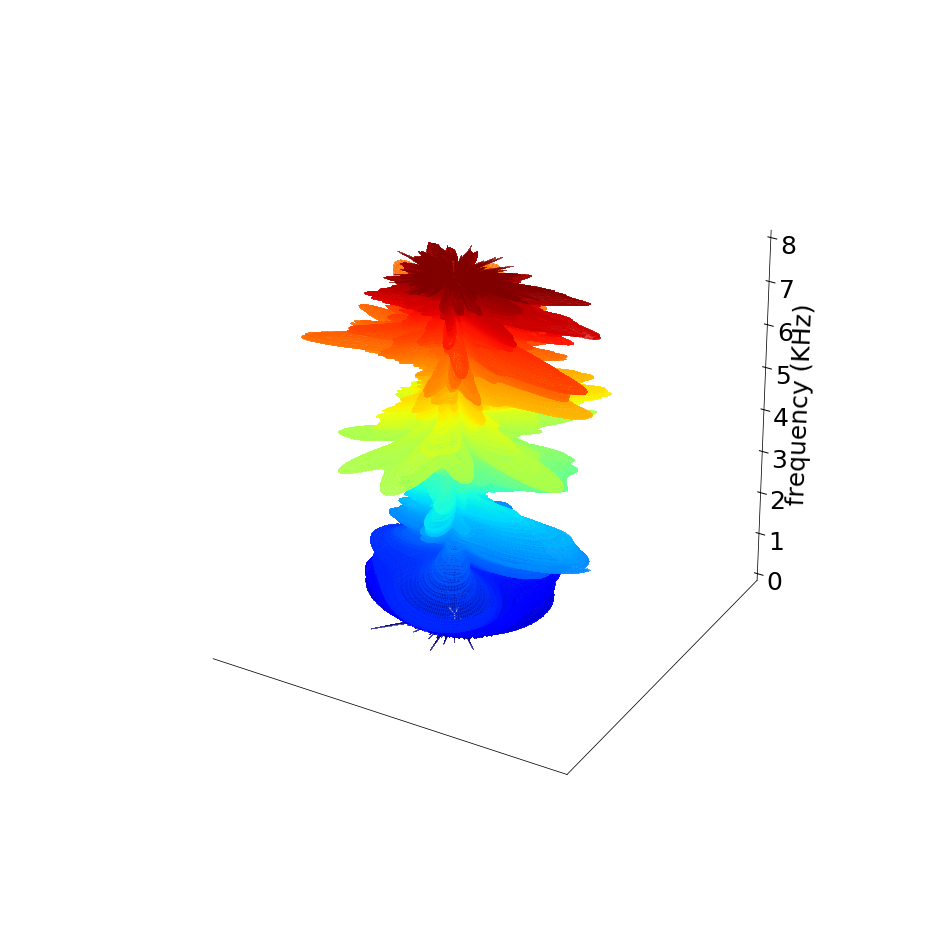}} 
  \centerline{(c) LEGO2}
\end{minipage}%
\begin{minipage}[b]{0.5\linewidth}
  \centering
  \centerline{\includegraphics[width=\linewidth]{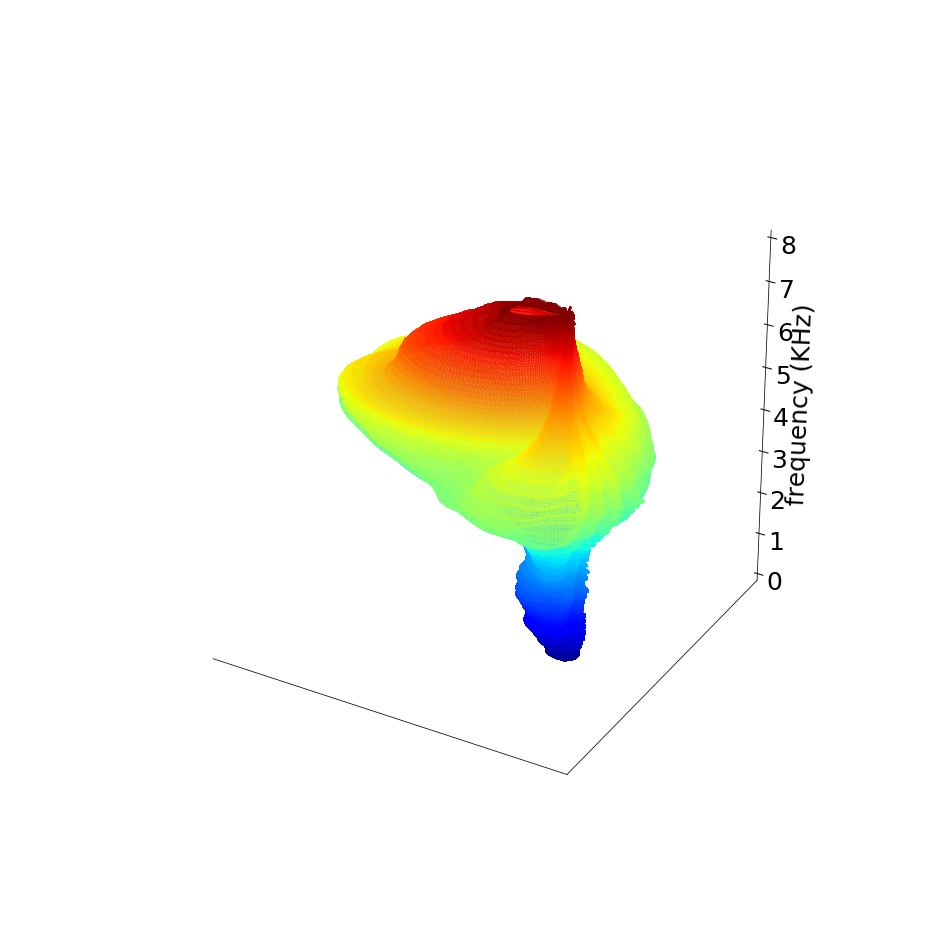}} 
  \centerline{(d) KEMAR}
\end{minipage}
\caption[]{Directional frequency magnitude response for different devices. Each horizontal slice is the polar pattern at the corresponding frequency between 0-8000 Hz from bottom to top. The colors only aid visualization.}
\label{fig:responses}
\end{figure}
At least conceptually, monaural localization is a simple matter if the source is always the same: for each direction the HRTF imprints a distinct spectral signature onto the sound which can be detected through correlation. In reality, the sources are diverse but this fixed-source case lets us develop a good intuition.
\subsection{Intuition}
\label{sec:intuition}
To see how scattering helps, suppose the sources are white and a set of $D$ directional transfer functions $\{\cH_d\}_{d=1}^{D}$ of our device is known. The power spectral density (PSD) of a white source is flat and scaled by the source's power:  $\mathbb{E}[|\cS_j|^{2}] = \sigma_j^2$. Assuming the noise has zero mean, the PSD of the observation is
\begin{equation}
\mathbb{E}[|\cY|^2] = \sum_{j \in \mathcal{J}} \sigma_j^2 |\cH_j|^2,
\end{equation}
which is a positive linear combination of the squared magnitudes of the transfer functions. In other words, $\mathbb{E}[|\cY|^2]$ belongs to a cone defined as 
\begin{equation}
C_\mathcal{J} = \{\mathbf{x}: \mathbf{x}= \sum_{j \in \mathcal{J}} c_j |\cH_j|^2, \; c_j>0\},
\end{equation}
Each configuration of sources $\mathcal{J}$ results in a different cone $C_\mathcal{J}$. For $D$ directions and $J$ white sources, there are $D \choose J $ possible cones which are known a priori since we assume knowing the scatterer. These cones reside in an $F$-dimensional space of direction-dependent spectral magnitude responses, $\mathbb{R}_{+}^{F}$, rather than the physical scatterer space $\mathbb{R}^3$. While the arrangement of cones in $\mathbb{R}_{+}^{F}$ is indeed determined by the geometry of the device in $\mathbb{R}^{3}$, the relation is complicated and nonlinear, namely it requires solving a boundary value problem for the Helmholtz equation at each frequency. 

Thus, we have $\mathbb{E}[|\cY|^2] \in \underset{\mathcal{J}}{\bigcup} ~C_\mathcal{J}$, and in theory, the localization problem becomes one of identifying the correct cone
\begin{equation}
  \widehat{\mathcal{J}} = \argmin_\mathcal{J} \text{dist}\left( \widehat{\mathbb{E}}[|\cY|^2], C_\mathcal{J} \right),
\end{equation}
where $\widehat{\mathbb{E}} \big[ |\cY|^2 \big]$ denotes the empirical estimate of the corresponding expectation from observed measurements. We discuss this further in the next section where we give the complete algorithm. 

Testing for cone membership results in correct localization when $C_{\mathcal{J}_1} = C_{\mathcal{J}_2}$ implies $\mathcal{J}_1 = \mathcal{J}_2$ (distinct direction sets span distinct cones)---a condition that is loosely speaking more likely to hold the more \emph{diverse} $\cH_j$ are. Examples of $|\cH_j|$ are illustrated in Figure \ref{fig:responses}. In particular, Figure \ref{fig:responses}(a) corresponds to an omnidirectional microphone with a flat frequency response and no scattering structure. In this case $C_{\mathcal{J}} = \{ \sigma^2 \bf{1} : \sigma \geq 0 \}$ and monaural localization is impossible. Figure \ref{fig:responses}(d) corresponds to an HRTF which features relatively smooth variations. Finally, Figures \ref{fig:responses}(b) and \ref{fig:responses}(c) correspond to our devices constructed using LEGO bricks whose responses have more fluctuating variations. In a nutshell, scattering induces a union-of-cones structure that enables us to localize white sources using a single sensor; stronger and more diverse scattering implies easier localization. 

\subsection{White Noise Localization}
In this section we describe a simple algorithm for localizing noise sources based on the intuition provided in the previous section\footnote{This algorithm appears in our previous conference publication \cite{elbadawy2017}.}. Our experiments with white noise localization will provide us with an ideal case baseline.

First, we need to replace the expected value $\mathbb{E}[|\cY|^2]$ by its empirical mean computed from $N$ time frames. For many types of sources this approximation will be accurate already with a small number of frames by the various concentration of measure results \cite{ledoux}; we corroborate this claim empirically.

Second, for simplicity, we replace each cone $C_\mathcal{J}$ by its smallest enclosing subspace $\mathcal{S}_\mathcal{J} = \mathrm{span} \left\{ |\cH_{j}|^2 \right\}_{j \in \mathcal{J}}$ represented by a matrix
\[
  \mathbf{B}_\mathcal{J} \bydef \left[ \ |\cH_{j_1}|^2, \ \ldots,\  |\cH_{j_J}|^2 \ \right], \ j_k \in \mathcal{J}.
\] 
This way the closest cone can be approximately determined by selecting $\mathcal{J} \subseteq \mathcal{D}$ such that the subspace projection error is the smallest possible. The details of the resulting algorithm are given in Algorithm \ref{alg:exh}; note the implicit assumption that $J < F$ as otherwise all cones lie in the same subspace. 

\begin{algorithm}[t]
\caption{White Noise Localization}
\begin{algorithmic}
 \REQUIRE Number of sources $J$, magnitudes of directional transfer functions $\{|\cH_j|^2\}_{j \in \mathcal{D}}$, $N$ audio frames $\cY \in \mathbb{C}^{F\times N}$.
    \ENSURE Directions of arrival $\widehat{\Theta} = \{\widehat{\theta}_1,\dots , \widehat\theta_J\}$.
    \STATE Compute the empirical PSD ${\mathbf{y}} = \frac{1}{N}\sum_{n=1}^{N}|\cY_n|^{2} $
    \FOR{\text{every} $\mathcal{J} \subseteq \mathcal{D}$, $|\mathcal{J}| = J$}
      \STATE $\mathbf{B}_\mathcal{J} \gets \left[ |\cH_j|^2 \right]_{j \in \mathcal{J}}$
      \STATE $\mathbf{P}_\mathcal{J} \gets \mathbf{B}_\mathcal{J} \mathbf{B}_\mathcal{J}^\dag$
    \ENDFOR
    \STATE $\widehat{\mathcal{J}} \gets \underset{\mathcal{J} }{\mathrm{arg\,min}}\;\| (\mathbf{I} - \mathbf{P}_\mathcal{J}) {\mathbf{y}} \|$
    \STATE $\widehat{\Theta} \gets \left\{ \theta_j \ | \ j \in \widehat{\mathcal{J}} \right\}$
\end{algorithmic}
\label{alg:exh}
\end{algorithm}

The robustness of Algorithm \ref{alg:exh} to noise largely depends on the angles between pairs of subspaces $\mathcal{S}_\mathcal{J}$ for different configurations $\mathcal{J}$, with smaller angles implying a higher likelihood of error. Intuitively, a transfer function that varies smoothly across directions is unfavorable as it yields smaller subspace angles (more similar subspaces).

We now turn our attention to the realistic case where sound sources are diverse: how can we determine whether an observed spectral variation is due to the directivity of the sensor or a property of the sound source itself? In fact, localization of unfamiliar sounds degrades not only for monaural but also binaural listening \cite{hebrank1974}. It has also been found that older children with unilateral hearing loss perform better in localization tasks than younger children \cite{uhlchildren}. We can thus conclude that both knowledge and experience allow us to dissociate source spectra from directional cues. Once the HRTF and the source spectra have been learned, it becomes possible to differentiate directions based on their modifications by the scatterer. 
\section{Method}
\label{sec:main}
We can think of an ideal white source as belonging to the subspace $\textrm{span}\{\mathbf{1}\}$ since $|\cS|^{2} = \mathbf{1}\sigma^2 $. In the following, we generalize the source model to more interesting signals such as speech. For those signals, testing for cone membership the same way we did for white sources is not straightforward. We can, however, take advantage of the non-negativity of the data to design efficient localization algorithms based on NMF. Instead of continuing to work with power spectra $|\cS|^{2}$, we switch to magnitude spectra $|\cS|$: prior work \cite{virtanen2007,mohammadiha2013} and our own experiments found that magnitude spectra perform better in this context.

\subsection{Problem Statement}
\label{sub:statement}
We adopt the usual assumption that magnitude spectra are additive \cite{virtanen2007,smaragdis2007}. Then the magnitude spectrogram of the observation \eqref{eq:mixspec} can be expressed as
\begin{equation}
\mY = \sum_{j \in \mathcal{J}} \textrm{diag}(\mH_j) \mS_j + \mathbf{E},
\label{eq:magspec}
\end{equation}
for $\mY = |\cY|$, $\mH = |\cH|$, $\mS_j = |\cS_j|$, and $\mathbf{E} = |\mathsf{E}|$. We further model the source $\mS_j$ as a non-negative linear combination of $K$ atoms $\mW\in \mathbb{R}_{+}^{F\times K}$ such that $\mS_j = \mW \mX_j$. The atoms in $\mW$ can correspond to either spectral prototypes of the sources to be localized or they can be learned from training data. Using this source model, we rewrite \eqref{eq:magspec} as
\begin{equation}
\mY = \mA \mX + \mathbf{E},
\label{eq:mix}
\end{equation}
where $\mY \in \mathbb{R}_{+}^{F \times N}$ is the observation, $$\mA = \big[ \, \textrm{diag}(\mH_1)\mW, \dots, \textrm{diag}(\mH_D)\mW \, \big] \in \mathbb{R}_{+}^{F\times KD }$$ is the mixing matrix, and $$\mX  = \big[ \, \mX_1^\mathrm{T}, \dots, \mX_D^\mathrm{T} \, \big]^{\mathrm{T}} \in \mathbb{R}_{+}^{KD \times N}$$ are the dictionary coefficients. Each group $\mX_d \in \mathbb{R}_{+}^{K\times N}$ corresponds to the set of coefficients for one source at one direction $d$. 

For localization, we wish to recover $\mX$; however, we are not interested in the coefficient values themselves but rather whether given coefficients are active or not---the activity of a coefficient indicates the presence of a source. In other words, we are only concerned with identifying the support of $\mX$. Localization is achieved by selecting the $J$ directions whose corresponding groups $\mX_d$ have the highest norms. 

\subsection{Regularization}
Still, recovering $\mX$ from \eqref{eq:mix} is an ill-posed problem. To get a reasonable solution, we must regularize by prior knowledge about $\mX$. We thus make the following two assumptions. First, the sources are few ($J\ll D$), which means that most groups $\mX_d$ are zero. Second, each source has a sparse representation in the dictionary $\mW$. These assumptions are enforced by considering the solution to the following penalized optimization problem
\begin{equation}
\underset{\mX \geq 0}{\rm arg\,min}\; D(\mY \, \| \, \mA \mX) + \lambda \Psi_g (\mX) + \gamma \Psi_s (\mX),
\label{eq:cost}
\end{equation}
where $D(\cdot \, \| \, \cdot)$ is the data fitting term, $\Psi_g$ is a group-sparsity penalty to enforce the first assumption, and $\Psi_s$ is a sparsity penalty to enforce the second assumption. The parameters $\lambda>0$ and $\gamma>0$ are the weights given to the respective penalties.

A common choice of $D(\cdot \, \| \, \cdot)$ for speech is the Itakura--Saito divergence \cite{fevotte2009}, which for strictly positive scalars $v$ and $\hat{v}$, is defined as
\begin{equation}
d_{IS}(v \, \| \, \hat{v})= \frac{v}{\hat{v}} - \log \frac{v}{\hat{v}} - 1,
\end{equation}
so that $D(\mathbf{V} \, \| \, \hat{\mathbf{V}}) = \sum_{fn} d_{IS}(v_{fn}||\hat{v}_{fn})$. Another option is the Euclidean distance 
\begin{equation}
D(\mathbf{V} \, \| \, \hat{\mathbf{V}}) =\frac{1}{2} \sum_{fn}( v_{fn}-\hat{v}_{fn})^2.
\end{equation}
Both the Itakura--Saito divergence and the Euclidean distance belong to the family of $\beta$-divergences with $\beta=0$ and $\beta=2$ respectively \cite{fevotte2011}. The former is scale-invariant and is thus preferred for audio which has a large dynamic range \cite{fevotte2009}. 

To promote group sparsity, we choose $\Psi_g$ to be the $\log/\ell_1$ penalty \cite{lefevre2011} defined as 
\begin{equation}
\Psi_g (\mX) = \sum_{d=1}^{D} \log (\epsilon + \|\textrm{vec}(\mX_d)\|_1),
\end{equation}
where $\textrm{vec}(\cdot)$ is a vectorization operator.
To promote sparsity of the dictionary expansion coefficients, we choose $\Psi_s$ to be $\ell_1$-norm \cite{donoho2004l1} as 
\begin{equation}
\Psi_s (\mX) = \|\textrm{vec}(\mX)\|_1.
\end{equation}
The combination of sparsity and group-sparsity penalties results in a small number of active groups that are themselves sparse. Thus the joint penalty is known as sparse-group sparsity \cite{spgl}. 

We note that our main optimization \eqref{eq:cost} is performed only over the latent variables $\mathbf{X}$; the non-negative dictionary $\mA$, which is constructed by merging a source dictionary learned by off-the-shelf implementations of standard algorithms with the direction-dependent transfer functions as described in Section \ref{sub:statement}, is taken as input. We thus avoid the joint optimization over $\mA$ and $\mX$ which is a major source of non-convexity. However, our choices for non-convex functionals like the Itakura-Saito divergence and the $\log / \ell_1$ penalty (although the latter is quasi-convex) render the whole optimization \eqref{eq:cost} non-convex.
\subsection{Derivation}
The minimization \eqref{eq:cost} can be solved iteratively by multiplicative updates (MU) which preserve non-negativity when the variables are initialized with non-negative values. 
The update rules for $\mX$ are derived using maximization-minimization for the group-sparsity penalty in \cite{lefevre2011} and for the $\ell_1$-penalty in \cite{fevotte2011}. They amount to dividing the negative part of the gradient by the positive part and raising to an exponent. In the following we derive the MU rules for our objective \eqref{eq:cost}. 

Note that the objective is separable over the columns of $\mX$
\begin{equation}
C(\vx) \ = \ D(\vy \, \| \,  \mA \vx) + \lambda \sum_{d=1}^{D} \log (\epsilon + \|\vx_d\|_1) + \gamma \|\vx\|_1,
\label{eq:costc}
\end{equation}
where $\vy\in \mathbb{R}_{+}^{F}$, $\vx\in \mathbb{R}_{+}^{FK}$ are columns of $\mY$ and $\mX$ respectively. With $\vx^{(i)}$ as the current iterate, the gradient of \eqref{eq:costc} with respect to one element $x_k$ of $\vx$ when $D(\cdot \, \| \, \cdot)$ is the Itakura--Saito divergence is given by
\begin{align}
\nabla_{x_k} C(\vx^{(i)}) = &- \sum_f  y_f (\mA \vx^{(i)})_f^{-2} a_{fk} \nonumber \\&+  \sum_f   (\mA \vx^{(i)})_f^{-1} a_{fk} 
+ \lambda \frac{1}{\epsilon+\|\vx^{(i)}_d\|_1} + \gamma,
\end{align}

where $a_{fk} = [\mA]_{fk}$ are entries of $\mA$.  The update rule is then given as
\begin{align}
x^{(i+1)}_k 
\ &= \ x^{(i)}_k \left(\frac{\nabla_{x_k}^{-} C(\vx^{(i)})}{\nabla_{x_k}^{+} C(\vx^{(i)})}\right)^{\frac{1}{2}} \nonumber \\
\ &= \ x^{(i)}_k \left(\frac{ \sum_f  y_f (\mA \vx^{(i)})_f^{-2}a_{fk} }{ \sum_f ( \mA \vx^{(i)})_f^{-1} a_{fk}  + \lambda \frac{1}{\epsilon+\|\vx^{(i)}_d\|_1} + \gamma}\right)^{\frac{1}{2}},
\end{align}
where $\frac{1}{2}$ is a corrective exponent \cite{fevotte2011}. The updates in matrix form are shown in Algorithm \ref{alg:spgnmf} where the multiplication $\odot$, division, and power operations are elementwise and $\mathbf{P}$ is a matrix of the same size as $\mX$. Also shown are the updates for using the Euclidean distance following \cite{fevotte2011,cichocki2006} where $[v]_\epsilon = \textrm{max}\{v,\epsilon\}$ is a thresholding operator to maintain non-negativity with $\epsilon=10^{-20}$. 

\begin{algorithm}
\caption{MU for NMF with Sparse-group Sparsity}
\label{alg:spgnmf}
\begin{algorithmic}
\REQUIRE $\mY$, $\mathbf{A}$, $\lambda$, $\gamma$
\ENSURE $\mX$
\STATE Initialize $\mX = \mathbf{A}^{\textrm{T}}\mY$
\STATE $\mathbf{\widehat{Y}} \gets \mathbf{AX}$
\REPEAT
\FOR{$d=1,\dots,D$}
\STATE $\mathbf{P}_{d} \gets \dfrac{1}{\epsilon + \|{\textrm{vec}(\mathbf{X}_{d})}\|_1}$
\ENDFOR
\IF{Itakura--Saito}
\STATE $\mX \gets \mX \odot \left(\dfrac{\mathbf{A}^T (\mathbf{Y}\odot \mathbf{\widehat{Y}}^{-2})}{\mathbf{A}^T\mathbf{\widehat{Y}}^{-1}+\;  \lambda \mathbf{P}+\; \gamma}\right)^{\frac{1}{2}} $
\vspace{1mm}
\ELSIF{Euclidean} \vspace{2mm}
\STATE $\mX \gets \mX \odot \left[\dfrac{\mathbf{A}^T \mathbf{Y}- \;  \lambda \mathbf{P}-\; \gamma}{\mathbf{A}^T\mathbf{\widehat{Y}}}\right]_\epsilon $
\ENDIF
\STATE $\mathbf{\widehat{Y}} \gets \mathbf{AX}$
\UNTIL{convergence}
\end{algorithmic}
\end{algorithm}
\subsection{Algorithm}
\label{sec:alg}
The discretization of the azimuth into $D$ evenly-spaced directions has a direct correspondence with the localization errors. On the one hand, a course discretization limits the localization accuracy to approximately the size of the discretization bin $\frac{360}{D}^{\circ}$. On the other hand a fine discretization may warrant a smaller error floor, but it implies a model matrix with a higher coherence only worsening the ill-posedness of the optimization problem \eqref{eq:cost}. It additionally results in a larger matrix which hampers the matrix factorization algorithms that are of complexity $\mathcal{O}(FKDN)$ per iteration \cite{fevotte2009,lefevre2011}. A common compromise is the multiresolution approach \cite{malioutov2005,saxena2009} in which position estimates are first computed on a coarse grid, and then subsequently refined on a finer grid concentrated around the initial guesses. We test the following strategy:
\begin{enumerate}
\item Attempt localization on a coarse grid,
\item Identify the top $T$ direction candidates,
\item Construct the model matrix using the $T$ candidates and their neighbors at a finer resolution,
\item Rerun the NMF localization.
\end{enumerate} 
The final algorithm for source localization by NMF with and without multiresolution is shown in Algorithm \ref{alg:sourceloc}. Since \eqref{eq:cost} is non-convex, different initializations of $\mathbf{X}$ might lead to different results. We thus later run an experiment to test the influence on the actual localization performance in Section \ref{sec:exp}.

\begin{algorithm}
\caption{Direction of Arrival Estimation by NMF}
\label{alg:sourceloc}
\begin{algorithmic}
\REQUIRE Observation $y(t)$, Number of sources $J$, Parameter for group sparsity $\lambda$, Parameter for $\ell_1$ sparsity $\gamma$, magnitudes of directional transfer functions $\{\mH_j\}_{j \in \mathcal{D}}$, source model $\mW$
\ENSURE Directions of arrival $\widehat{\Theta} = \{\widehat{\theta}_1,\dots , \widehat\theta_J\}$
\STATE Construct $\mA \gets \big[ \, \textrm{diag}(\mH_1)\mW, \dots, \textrm{diag}(\mH_D)\mW \, \big]$
\STATE Construct $\mY \gets |\textrm{STFT}\{y\}|$
\STATE Factorize $\mY\approx\mA\mX$ using Algorithm \ref{alg:spgnmf}
\STATE Calculate $\mathcal{D} = \{\|\textrm{vec}(\mX_d)\|_1$ for $d=1,2,\dots, D\}$
\IF{Multiresolution}
\STATE Identify $T$ candidates and their $RT$ neighbors $\{\mH_{t,r}\}_{t=1,r=0}^{t=T,r=R}$
\STATE Construct $\widetilde{\mA} \gets \big[\textrm{diag}(\mH_{1,0})\mW, \dots, \textrm{diag}(\mH_{T,R})\mW\big]$
\STATE Factorize $\mY\approx\widetilde{\mA}\widetilde{\mX}$ using Algorithm \ref{alg:spgnmf}
\STATE Calculate $\mathcal{D} = \{\|\textrm{vec}(\widetilde{\mX}_d)\|_1$ for $d=1,2,\dots, (R+1)T\}$
\ENDIF 
\STATE $\widehat{\mathcal{J}} \gets \{$Indices of the $J$ largest elements in $\mathcal{D}\}$
\STATE $\widehat{\Theta} \gets \left\{ \theta_j \ | \ j \in \widehat{\mathcal{J}} \right\}$
\end{algorithmic}
\end{algorithm}

\section{Experimental Results}
\label{sec:exp}
\subsection{Devices}
\begin{figure*}
\hfill
\begin{minipage}[b]{0.3\linewidth}
  \centering
  \centerline{\includegraphics[width=0.9\linewidth,height=0.7\linewidth]{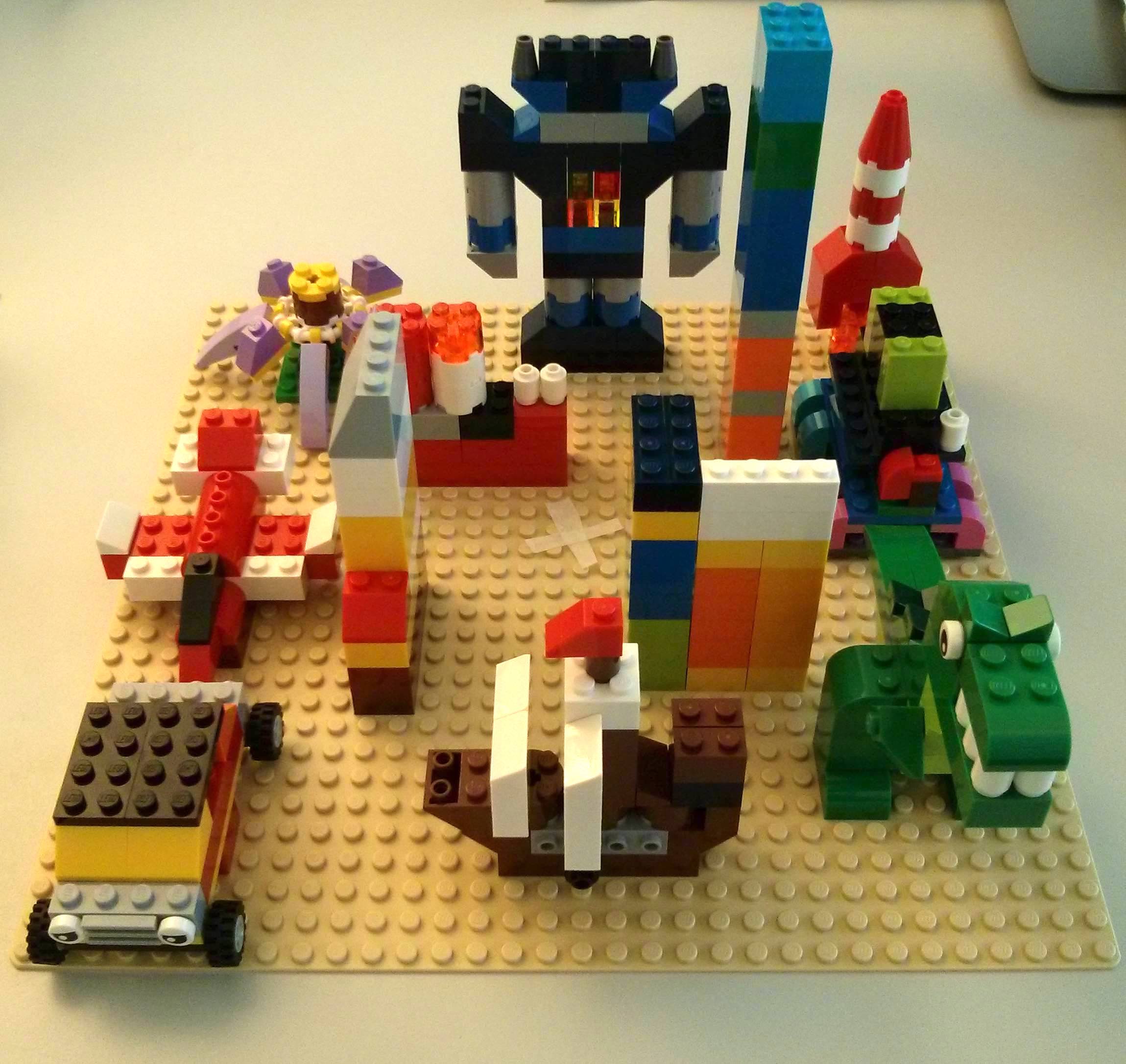}}
  \centerline{(a)}\medskip
\end{minipage}
\begin{minipage}[b]{0.3\linewidth}
  \centering
  \centerline{\includegraphics[width=0.9\linewidth,height=0.7\linewidth]{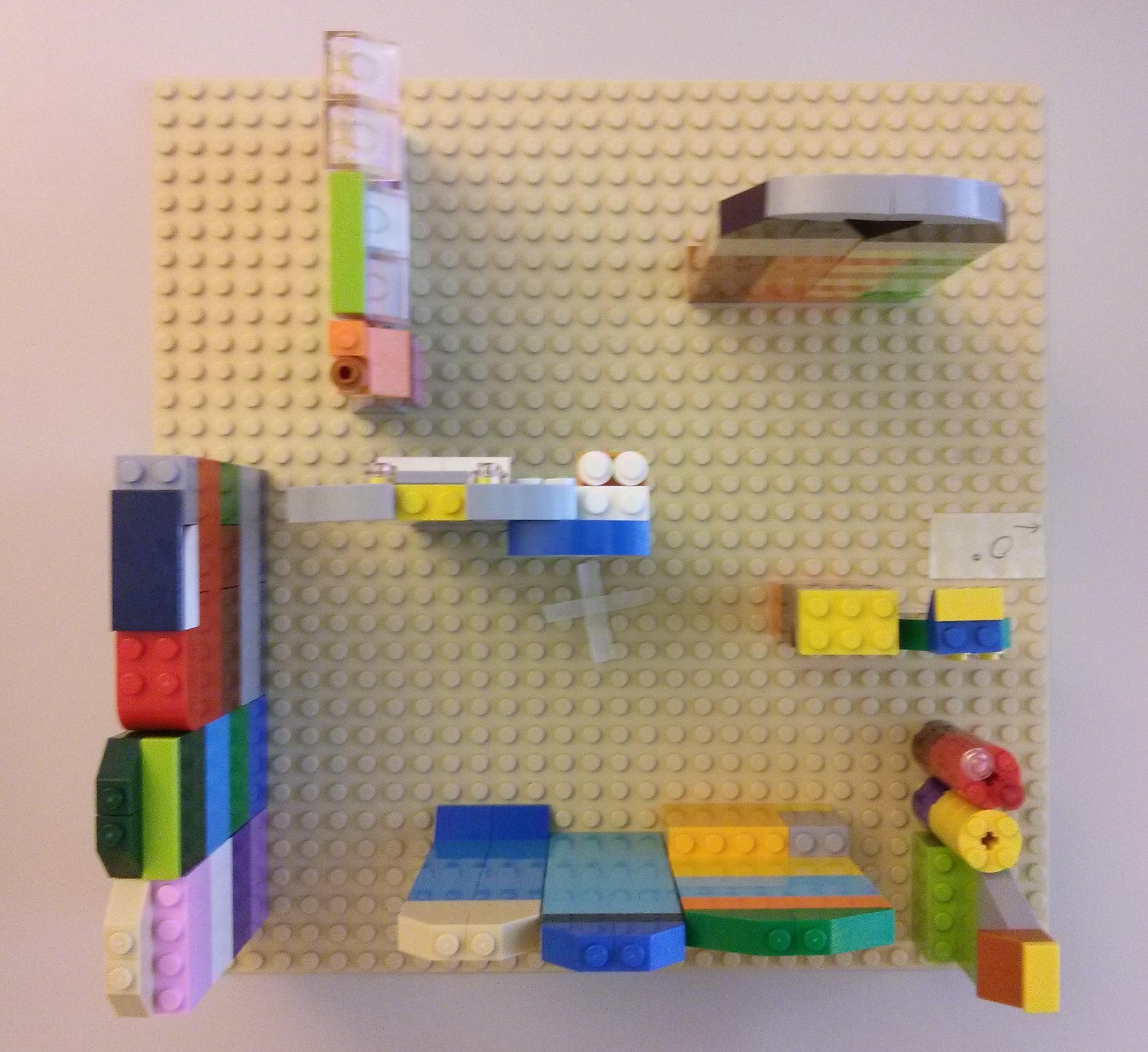}} 
  \centerline{(b)}\medskip
\end{minipage}
\begin{minipage}[b]{0.3\linewidth}
  \centering
  \centerline{\includegraphics[width=0.9\linewidth,height=0.7\linewidth]{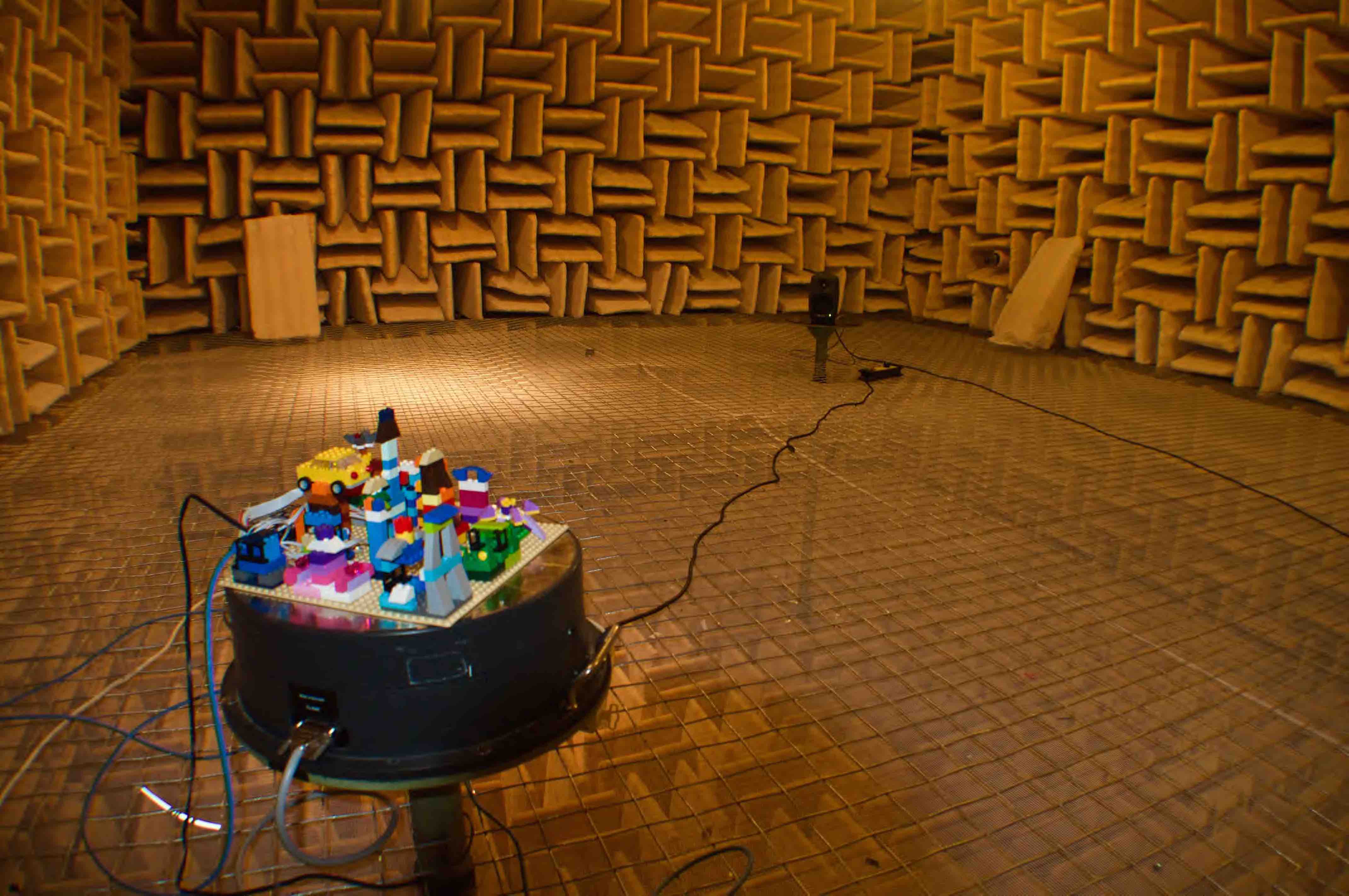}} 
  \centerline{(c)}\medskip
\end{minipage}
\hfill
\caption[]{Sensing devices made of LEGO bricks. The location of the microphone is marked by an ``x". (a) LEGO1. (b) LEGO2. (c) Calibration setup in an anechoic chamber.} 
\label{fig:lego}
\hfill
\end{figure*}

We ran experiments using three different devices:

\paragraph{LEGO1 and LEGO2} The first two devices are structures composed of LEGO bricks as shown in Figure \ref{fig:lego}. Since we aimed for diverse random-like scattering, we stacked haphazard brick constructions on a base plate of size 25 cm $\times$ 25 cm along with one omnidirectional microphone. The heights of the different constructions vary between 4 and 12.5 cm. We did not attempt to optimize the layout. The only assumption we make regarding the dimensions of the device is that some energy of the target source resides at frequencies where the device observably interacts with the acoustic wave. We note that the problem of designing and optimizing the structure to get a desired response is that of inverse obstacle scattering which is a hard inverse problem in its own right \cite{colton, colton2000}. For the present work, we simply observe that our random structures result in the desired random-like scattering.  

The directional impulse response measurements were then done in an anechoic chamber where the device was placed on a turntable as shown in Figure \ref{fig:lego}(c) and a loudspeaker at a distance of 3.5 m emitted a linear sweep. We note that the turntable is symmetric, so its effect on localization in the horizontal plane, if any, is negligible. The duration of the measured impulse responses averages around 20 ms. Figures \ref{fig:responses}(b) and \ref{fig:responses}(c) show the corresponding magnitude response for the two devices. Due to their relatively small size, they mostly scatter high frequency waves and so the response at lower frequencies is comparably flat. We thus expect that only sources with enough energy in the higher range of frequencies can be accurately localized. 

\paragraph{KEMAR} The third device is KEMAR \cite{kemarhrtf} which is modeled after a human head and torso so that its response accurately approximates a human HRTF. The mannequin's torso measures $44 \times 24 \times 73$ cm and the head's diameter is 18 cm. The duration of the impulse response is 10 ms. Figure \ref{fig:responses}(d) shows the corresponding magnitude response. As can be seen, the variation across the directions is very smooth which we expect to result in worse monaural localization performance. 

\subsection{Data and parameters}
The mixtures are created by first convolving the source signals with the impulse responses and then corrupting the result by additive white Gaussian noise at various levels of signal-to-noise ratio defined as $$\text{SNR} = 20 \log \frac{\|\sum_j s_j(t) \ast h_j(t)\|_2}{\|e(t)\|_2}~\text{dB}.$$ We use frame-based processing using the STFT with a Hann window of length 64 ms, with a 50\% overlap. The number of iterations in NMF (Algorithm \ref{alg:spgnmf}) was set to 100. 

The test data contains 10 speech sources (5 female, 5 male) from TIMIT \cite{timit} sampled at 16000 Hz. The duration of the speech varies between 3.1 and 4.5 s and the maximum amplitude is normalized to 1 so that all sources have the same volume. No preprocessing of the sources such as silence removal was done; when mixing two sources, the longest one was truncated.

A separate validation set was used to select the best sparsity parameters for each device. The parameters that gave the best performance averaged for one and two sources were chosen. We additionally tested whether the lower frequencies can be ignored in localization since, as mentioned before, for the relatively small scatterers the lower frequency range lacks variation and is thus uninformative. Moreover, truncating the lower frequencies would help reduce coherence between the directional transfer functions. The final parameters and used frequency range are summarized in Table \ref{table:params}.   
\begin{table}[t]
\caption{Parameters per device.}
\centering
\begin{tabular}{@{}llll@{}}
\toprule
 & LEGO1 & LEGO2 & KEMAR \\\midrule
Frequency & 3000-8000 Hz & 3000-8000 Hz & 0-8000 Hz\\
Prototypes & $\lambda = 10$, $\gamma=10$  & $\lambda = 10$, $\gamma=1$  &  $\lambda = 10$, $\gamma=0.1$ \\
USM ($\beta=0$) & $\lambda = 0.1$, $\gamma=10$   & $\lambda = 10$, $\gamma=1$  & $\lambda = 100$, $\gamma=10$    \\
USM ($\beta=2$) & $\lambda = 1$, $\gamma=1$  & $\lambda = 1$, $\gamma=1$  &  $\lambda = 1$, $\gamma=1$ \\
Multiresolution & $\lambda = 0.1$, $\gamma=1$  & $\lambda = 100$, $\gamma=0.1$  &  - \\
\bottomrule
\end{tabular}
\label{table:params}
\end{table}
\paragraph*{Source Dictionary}
For speech localization, we test two source dictionaries. For the first experiment, we use a dictionary of prototypes of magnitude spectra from 4 speakers (2 female, 2 male) in the test set.

For the second experiment, we use a more general universal speech model (USM) \cite{usm} learned from a training set of 25 female and 25 male speakers, also from TIMIT. We use a random initialization for the NMF when learning the USM. Each speaker in the training set is modeled using $K=10$ atoms, thus the final USM is $\mW \in \mathbb{R}_{+}^{F\times 500}$. In total, we use four versions of the USM in the experiments. Two versions correspond to learning the model by minimizing either the Itakura--Saito divergence or the Euclidean distance. The other two versions correspond to learning the model using only the subset of frequencies to be utilized in the localization. 

\subsection{Evaluation}
We estimate the azimuth of the sources in the range $[0^\circ,360^\circ)$. The model \eqref{eq:mix} assumes a discrete set of 36 evenly spaced directions while the sources are randomly placed on a finer grid of 360 directions. Given the estimated directions $\hat{\Theta}= \{\hat{\theta}_1,\dots, \hat{\theta}_J\}$ and the true directions $\Theta=\{\theta_1,\dots, \theta_J\}$, the localization error is computed as the average absolute difference modulo $360^\circ$ as
\begin{equation}
\underset{\mathcal{\pi}}{\textrm{min}}\; \frac{1}{J} \sum_{j \in \mathcal{J}} \left| (\hat{\theta}_{\mathcal{\pi}(j)} - \theta_j + 180) \;\textrm{mod} \; 360 - 180\right|,
\end{equation}
where $\mathcal{\pi}: \mathcal{J} \to \mathcal{J}$ is a permutation that best matches the ordering in $\hat{\Theta}$ and $\Theta$.

For each experiment, we test 5000 random sets of directions. We emphasize that we have been careful to avoid an inverse crime, and we produced the measurements by convolution in the time domain, not by multiplication in the STFT domain. Thus in this set up, the reported errors also reflect the modeling mismatch. 

Following \cite{woodruff2012}, we report the \emph{accuracy} defined as the percentage of sources localized to their closest $10^\circ$-wide bin as well as the mean error for those accurately localized sources. For 36 bins, there is an inherent average error of $2.5^\circ$. Thus, ideally the accuracy would be 100\% and the error $2.5^\circ$. Additionally, we report the accuracy per source, that is, the rate at which a source is correctly localized regardless of the other sources.

\subsection{NMF Initialization}
 Since in a non-convex problem different initializations might lead to different results, we run an experiment to test the effect of the initialization of $\mathbf{X}$ on the localization performance. The experiment consists of 300 tests for localizing one female speaker using LEGO2 and a USM. We compare the initialization mentioned in Algorithm \ref{alg:spgnmf} ($\mathbf{X} = \mathbf{A}^{\mathrm{T}}\mathbf{Y}$)\footnote{We use a deterministic initialization to facilitate reproducibility and multithreaded implementations.} to different random initializations. The estimated DoAs were in agreement for both initializations 98.67\% of the time with Itakura-Saito and 97\% with Euclidean distance. We show in Table \ref{table:acc} the localization accuracy rates for that experiment which are comparable. This means that there are either ``hard'' situations where localization fails regardless of the initialization or ``easy'' situations where it succeeds regardless of the initialization. Certainly,  tailor-made initializations in the spirit of \cite{kitamura2016,langville2014} may work slightly better, but such constructions are outside the scope of this paper. Additionally, we note that in these works initializations are constructed for the basis matrix. In our case, this matrix is $\mA$ which is given as input to the algorithm.
\begin{table}[h]
\caption{Localization accuracy for different NMF initializations.}
\centering
\begin{tabular}{@{}lll@{}}
\toprule
& $\mathbf{A}^{\mathrm{T}}\mathbf{Y}$ &Random \\\midrule
 Itakura-Saito & 93.00\% & 93.33\% \\
 Euclidean &  89.67\% &  90.00\%\\
\bottomrule
\end{tabular}
\label{table:acc}
\end{table}

\subsection{White Noise Localization}
\label{app:white}
We first test the localization of one and two white sources at various levels of SNR using Algorithm \ref{alg:exh}. Each source is 0.5 s of white Gaussian noise. We compare the performance using the three devices LEGO1, LEGO2, and KEMAR described above. For white sources, using the full range of frequencies, not a subset, was found to perform better. 
 
The accuracy rate and the mean localization error for the different devices are shown in Table \ref{table:white10acc}. In the one source case, all devices perform well. The mean error achieved by the devices for one white source is close to the ideal grid-matched $2.5^\circ$ which is better than the reported $4.3^\circ$ and $8.8^\circ$ in \cite{saxena2009} using an HMM. For two sources, the accuracy of the LEGO devices is still high, though lower than for one source. At the same time the accuracy of KEMAR deteriorates considerably. This is consistent with the intuition that interesting scattering patterns such as those of the LEGO devices result in better localization. 

\begin{table*}[t]
\caption{Error for white noise localization at a discretization of $10^\circ$ }
\centering
\begin{tabular}{@{}llllllll@{}}
\toprule
& &  \multicolumn{2}{c}{LEGO1} & \multicolumn{2}{c}{LEGO2} & \multicolumn{2}{c}{KEMAR}  \\
& SNR & Accuracy & Mean & Accuracy & Mean & Accuracy & Mean\\\midrule
One source & 30 dB  & 99.56\% & $2.63^\circ$ & 96.64\% & $2.54^\circ$  & 92.06\%  & $2.72^\circ$\\
& 20 dB & 99.58\% & $2.63^\circ$ & 96.54\% & $2.53^\circ$ & 92.12\% & $2.71^\circ$  \\
& 10 dB & 99.60\% & $2.60^\circ$ & 96.42\% & $2.53^\circ$ & 91.78\% & $2.73^\circ$\\
Two sources & 30 dB & 94.72\% & $2.75^\circ$ & 83.64\% & $2.62^\circ$  & 25.22\%  & $3.44^\circ$\\
& 20 dB & 94.54\% & $2.75^\circ$ & 83.34\% & $2.62^\circ$ & 25.48\% & $3.45^\circ$ \\
& 10 dB & 92.32\% & $2.73^\circ$ & 81.52\% & $2.62^\circ$ & 21.20\% & $3.59^\circ$\\
\bottomrule
\end{tabular}
\label{table:white10acc}
\end{table*}

We also test the effect of the discretization on the localization performance. In Table \ref{table:whitecompareacc}, we report the localization errors using LEGO1 at three different resolutions: $2^\circ$, $5^\circ$, and $10^\circ$. We find that improving the resolution results in more accurate localization for both one and two sources but the average error is still larger than the ideal $0.5^\circ$ and $1.25^\circ$ for the $2^\circ$ and $5^\circ$ resolutions respectively, especially for two sources. Since white sources are flat, this observation highlights a limitation of the device itself in terms of coherent or ambiguous directions. 
\begin{table*}[t]
\caption{Discretization comparison for white noise localization using LEGO1.}
\centering
\begin{tabular}{llllllll}
\toprule
&  & \multicolumn{2}{c}{$2^\circ$} & \multicolumn{2}{c}{ $5^\circ$} &  \multicolumn{2}{c}{$10^\circ$}  \\
& SNR& Accuracy & Mean & Accuracy & Mean & Accuracy & Mean\\\midrule
One source & 30 dB  & 100.0\% & $0.52^\circ$ & 100.0\% & $1.27^\circ$  & 99.56\%  & $2.63^\circ$\\
& 20 dB & 100.0\% & $0.52^\circ$ & 100.0\% & $1.27^\circ$ & 99.58\% & $2.63^\circ$  \\
& 10 dB & 100.0\% & $0.54^\circ$ & 100.0\% & $1.26^\circ$ & 99.60\% & $2.60^\circ$\\
Two sources & 30 dB & 98.56\% & $0.70^\circ$ & 98.78\% & $1.43^\circ$  & 94.72\%  & $2.75^\circ$\\
& 20 dB & 98.50\% & $0.71^\circ$ & 98.70\% & $1.43^\circ$ & 94.54\% & $2.75^\circ$ \\
& 10 dB & 97.30\% & $0.82^\circ$ & 97.32\% & $1.47^\circ$ & 92.32\% & $2.73^\circ$\\
\bottomrule
\end{tabular}
\label{table:whitecompareacc}
\end{table*}

\subsection{Speech Localization with Prototypes}
We now turn to speech localization which is considerably more challenging than white noise, especially in the monaural setting. Using the three devices, we test the localization of one and two speakers at 30 dB SNR. 
In this first experiment, we use a subset of 4 speakers from the test data (two female, two male) and consider an easier scenario where we assume knowing the exact magnitude spectral prototypes of the sources. Still, localization with colored prototypes is harder compared to noise prototypes (as in \cite{metamaterials}). This scenario serves as a gauge for the quality of the sensing devices for localizing speech sources. We organize the results by the number of sources as well as by whether the speaker is male or female. We expect the localization of female speakers to be more accurate since they have relatively more energy in the higher frequency range where the device responses are more informative. 

The results for the three devices are shown in Table \ref{table:speechproto10acc}. As expected the overall localization performance by the less smooth LEGO scatterers is significantly better than by KEMAR. Also as expected, the localization of male speech is worse than female speech except for LEGO1. Similar to the white noise case, the accuracy for localizing two sources is lower in comparison to one source. Moreover, we find that the presence of one female speaker improves the accuracy for LEGO2 and KEMAR, most likely due to the spectral content.

\begin{table*}[t]
\caption{Error for speech localization using prototypes at a discretization of $10^\circ$ }
\centering
\begin{tabular}{@{}llllllllll@{}}
\toprule
&   \multicolumn{3}{c}{LEGO1} & \multicolumn{3}{c}{LEGO2} & \multicolumn{3}{c}{KEMAR}  \\
& Accuracy & Mean & Per Source & Accuracy & Mean & Per Source & Accuracy & Mean & Per Source\\\midrule
female speech & 98.48\% & $2.53^\circ$ & 98.48\%  & 96.94\% & $2.51^\circ$ & 96.94\% & 79.74\%  & $3.42^\circ$ & 79.74\%\\
male speech & 98.76\% & $2.56^\circ$ & 98.76\%& 96.00\% & $2.53^\circ$ & 96.00\% & 72.06\% & $3.35^\circ$ & 72.06\%\\
female/female & 75.24\% &$2.46^\circ$ &  87.07\%  & 78.28\% & $2.40^\circ$ & 88.31\% & 11.66\% & $3.50^\circ$ & 46.70\%\\
female/male & 76.60\%& $2.44^\circ$& 87.79\% & 74.36\% &$2.41^\circ$ & 86.17\% & 10.90\% &$3.59^\circ$& 44.47\%\\
male/male & 80.24\% & $2.43^\circ$& 89.82\% & 74.22\% & $2.39^\circ$ & 86.04\%   & 9.24\% &$3.91^\circ$& 43.09\%\\
\bottomrule
\end{tabular}
\label{table:speechproto10acc}
\end{table*}
\subsection{Speech Localization with USM}
In this experiment, we switch to a more realistic and challenging setup where we use a learned universal speech model. We compare the performance of the Itakura--Saito divergence to that of the Euclidean distance in the cost function \eqref{eq:cost}. The accuracy and mean error for the three devices are shown in Table \ref{table:speechdisc10acc}. We observe that using the Itakura--Saito divergence results in better performance in a majority of cases which is in line with the recommendations for using Itakura--Saito for audio.
\begin{table*}[t]
\caption{Error for speech localization using a USM at a discretization of $10^\circ$ }
\centering
\begin{tabular}{@{}llllllllll@{}}
\toprule
&   \multicolumn{3}{c}{LEGO1} & \multicolumn{3}{c}{LEGO2} & \multicolumn{3}{c}{KEMAR}  \\
& Accuracy & Mean & Per Source & Accuracy & Mean & Per Source & Accuracy & Mean & Per Source\\\midrule
\emph{Itakura--Saito}\\
female speech & 93.20\% & $2.67^\circ$ & 93.20\% & 93.72\% & $2.54^\circ$ & 93.72\% & 46.56\% & $3.33^\circ$ & 46.56\%\\
male speech & 89.80\% & $2.74^\circ$ & 89.80\%& 87.70\% & $2.66^\circ$& 87.70\% & 35.56\% & $3.46^\circ$ & 35.56\%\\
female/female &26.38\%  & $2.64^\circ$ & 54.65\% & 53.52\% & $2.42^\circ$ & 73.93\% & 7.60\% & $3.90^\circ$ & 35.29\% \\
female/male & 24.76\% & $2.77^\circ$ & 54.42\%& 49.22\% & $2.49^\circ$& 70.93\% & 7.40\% & $4.01^\circ$ & 35.56\%\\
male/male &  19.78\%  & $3.02^\circ$ & 50.61\%& 39.54\% & $2.63^\circ$& 65.45\% & 7.44\%  & $4.36^\circ$ & 33.76\% \\[2mm]
\emph{Euclidean}\\
female speech &  85.60\% &  $2.79^\circ$ & 85.60\% & 91.26\% &  $2.57^\circ$& 91.26\% & 29.26\% & $3.75^\circ$ & 29.26\%\\
male speech & 76.00\% &  $2.78^\circ$ & 76.00\% & 86.74\% &  $2.65^\circ$& 86.74\% & 23.24\% &  $3.78^\circ$ & 23.24\%\\
female/female& 29.34\%&  $2.88^\circ$ & 56.66\% & 46.86\% & $2.48^\circ$ &69.89\%& 4.62\%  & $4.40^\circ$ & 23.75\%\\
female/male & 30.62\% &  $2.88^\circ$& 57.55\%&  42.28\% & $2.58^\circ$&66.40\%& 3.36\%& $4.34^\circ$ & 21.19\% \\
male/male & 23.72\%  & $2.96^\circ$ & 52.67\% & 35.50\% & $2.74^\circ$& 62.71\%& 2.80\% & $3.97^\circ$ & 18.60\%\\
\bottomrule
\end{tabular}
\label{table:speechdisc10acc}
\end{table*}

Similar observations as in the previous experiment hold with the LEGO scatterers offering better localization than KEMAR. We find that localizing one female speaker is successful with 93\% accuracy. Compared to the use of prototypes, the source model is here speaker-independent and the test set is larger containing 10 speakers; however, the accuracy is still only lower by 3-5\%. We also note that the mean localization error is $2.5^\circ$ which is smaller than the reported $7.7^\circ$ in \cite{saxena2009} with an HMM though at a lower SNR of 18 dB.

\begin{figure}[t]
\centering
\begin{minipage}[b]{0.5\linewidth}
  \centering
  \centerline{\includegraphics[width=\linewidth]{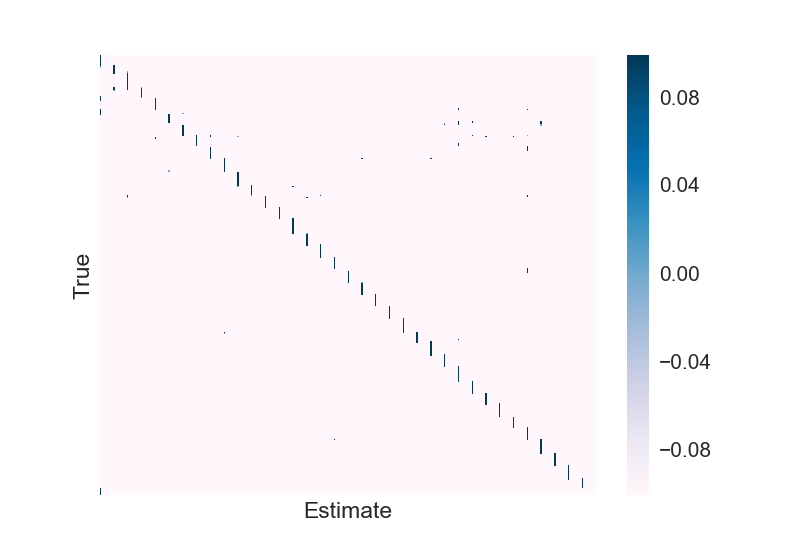}}
  \centerline{(a) $10^\circ$}\medskip
\end{minipage}%
\begin{minipage}[b]{0.5\linewidth}
  \centering
  \centerline{\includegraphics[width=\linewidth]{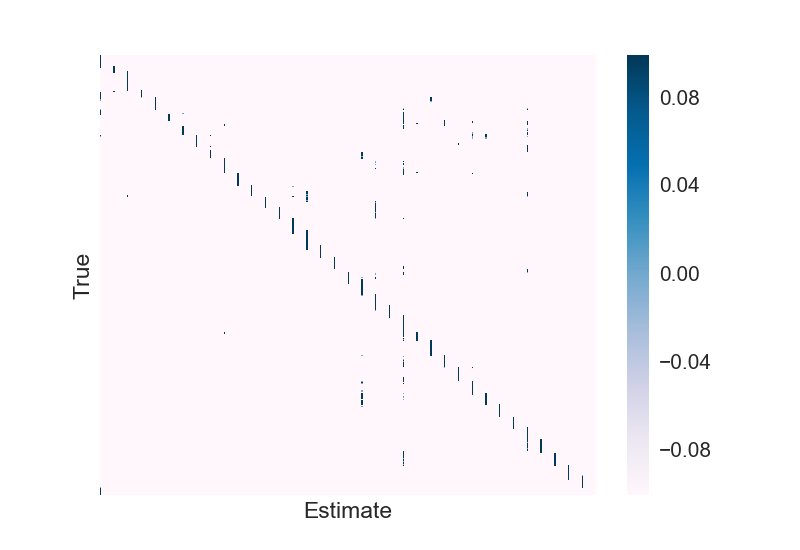}} 
  \centerline{(b) $10^\circ$}\medskip
\end{minipage}
\begin{minipage}[b]{0.5\linewidth}
  \centering
  \centerline{\includegraphics[width=\linewidth]{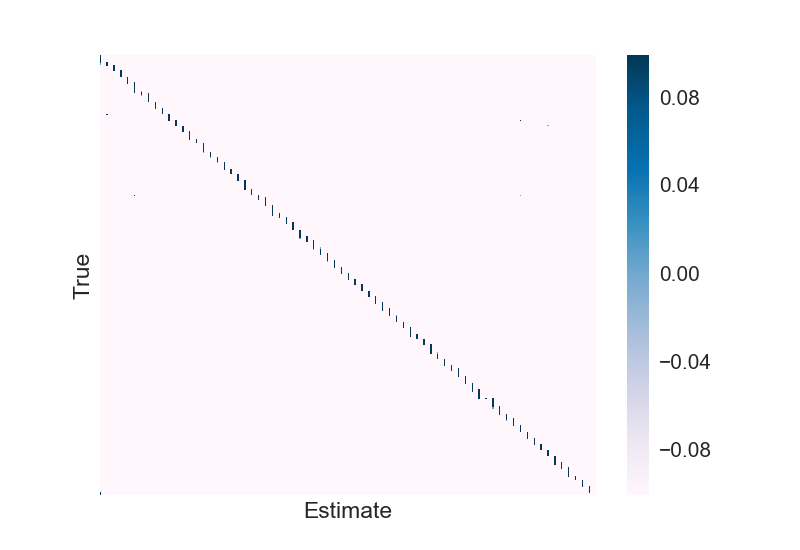}}
  \centerline{(c) $5^\circ$}\medskip
\end{minipage}%
\begin{minipage}[b]{0.5\linewidth}
  \centering
  \centerline{\includegraphics[width=\linewidth]{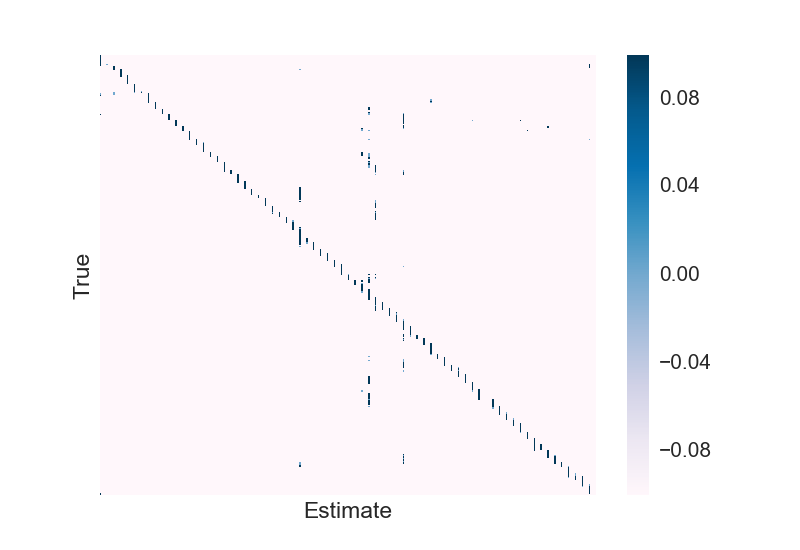}} 
  \centerline{(d) $5^\circ$}\medskip
\end{minipage}
\caption[]{Confusion matrices for localizing one speaker using LEGO2. Female speech has less outliers and improving the resolution decreases the number of outliers. Left: Female speech. Right: Male speech.} 
\label{fig:confmatrixm}
\end{figure}

\begin{figure*}[t]
\centering
\begin{minipage}[b]{0.3\linewidth}
  \centering
  \centerline{\includegraphics[width=\linewidth]{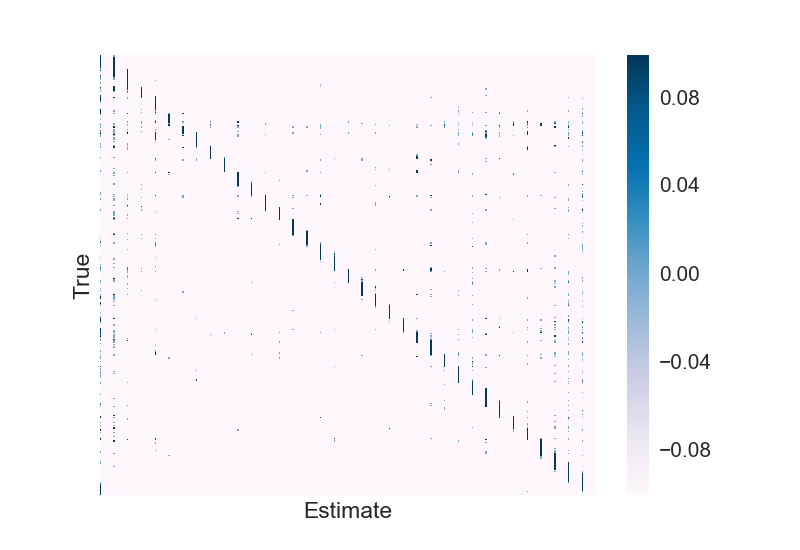}}
  \centerline{(a)}
\end{minipage}%
\begin{minipage}[b]{0.3\linewidth}
  \centering
  \centerline{\includegraphics[width=\linewidth]{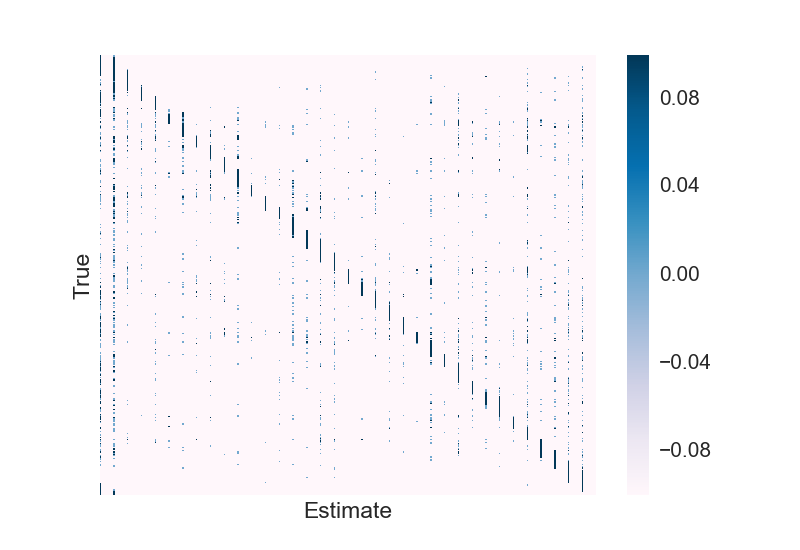}} 
  \centerline{(b)}
\end{minipage}%
\begin{minipage}[b]{0.3\linewidth}
  \centering
  \centerline{\includegraphics[width=\linewidth]{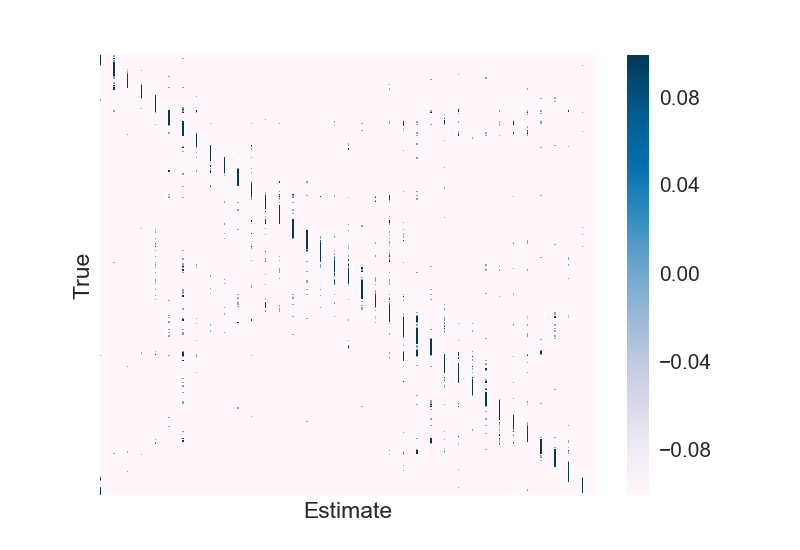}}
  \centerline{(c)}
\end{minipage}
\caption[]{Confusion matrices for localizing two sources using LEGO2 at a resolution of $10^\circ$. (a) With prototypes. (b) With a USM. (c) White sources.} 
\label{fig:confmatrix}
\end{figure*}

As expected, the localization accuracy for male speakers is lower than for female speakers. Since the mean errors are however not much larger than the ideal $2.5^\circ$, the lower accuracy points to the presence of outliers. We thus plot confusion matrices in Figures \ref{fig:confmatrix} and \ref{fig:confmatrixm} for female and male speakers respectively. On the horizontal axis, we have the estimated direction which is one of 36 only. First, we look at the single source case in Figures \ref{fig:confmatrixm}(a) and \ref{fig:confmatrix}(a) where we can clearly see the few outliers away from the diagonal. The number of outliers is larger for male speakers which is a direct result of the absence of spectral variation for male speech in the used higher frequency range.

For two sources, the number of outliers increases for both types as seen in Figure \ref{fig:confmatrixm}(b). We also plot in Figure \ref{fig:confmatrixm}(a) the confusion matrix for the case of using prototypes which has less outliers in comparison due to the stronger model. Note that outliers exist even with white sources as shown in Figure \ref{fig:confmatrixm}(c), which points to a deficiency of the device itself as mentioned before. However, we note that while the reported accuracy corresponds to correctly localizing the two sources simultaneously, the average accuracy per source which reflects the number of times at least one of the sources is correctly localized is often higher. For instance for female speakers, the accuracy is 53.52\% while the average accuracy per source is higher at 73.93\%. The overall best performance is achieved by LEGO2 with Itakura--Saito divergence. 

\subsubsection{Finer resolution}
As mentioned, one straightforward improvement to our system is to increase the resolution. We show in Table \ref{table:speechdisc5} the result of doubling the resolution from $10^\circ$ to $5^\circ$. For a single female speaker, the error is slightly higher than the ideal average of $1.25^\circ$ and the accuracy is improved relative to the initial bin size of $10^\circ$. While some improvement is apparent for the localization of one male speaker as well, the mismatch between the useful scattering range and source spectrum still prevents good performance. However, in line with the discussion in Section \ref{sec:alg}, localization of two sources is worse than at a coarser grid due to the increased matrix coherence, with the accuracy dropping from 55\% to 45\% for two female speakers.

\begin{table*}[t]
\caption{Error for speech localization at a resolution of $5^\circ$. }
\centering
\begin{tabular}{@{}llllllll@{}}
\toprule
&   \multicolumn{3}{c}{LEGO1} & \multicolumn{3}{c}{LEGO2}\\
& Accuracy & Mean & Per Source& Accuracy & Mean& Per Source\\\midrule
female speech & 97.08\% & $1.59^\circ$&  97.08\%&  99.72\% & $1.41^\circ$& 99.72\%   \\
male speech & 93.26\% & $1.76^\circ$ & 93.26\% & 92.68\% & $1.57^\circ$ & 92.68\%\\
female/female & 22.24\%& $1.95^\circ$& 55.25\% & 43.26\%& $1.47^\circ$ &71.23\%   \\
female/male & 21.60\%&  $2.14^\circ$& 55.33\%& 39.66\% &  $1.61^\circ$ & 68.82\% \\
male/male & 15.42\% &  $2.47^\circ$& 50.38\% & 29.72\%& $1.87^\circ$ & 63.31\% \\
\bottomrule
\end{tabular}
\label{table:speechdisc5}
\end{table*}

\subsubsection{Multiresolution}
Next we tested the multiresolution strategy where we refine the top estimates on the coarse grid using a search on a finer grid. We arbitrarily use the best 7 candidates at the $10^\circ$ grid spacing, and redo the localization at a finer $2^\circ$ grid centered around the 7 initial guesses. The hyperparameters for localization on the finer grid were tuned on a separate validation set and are given in Table \ref{table:params}.

\begin{figure}[t]
\centering
\begin{minipage}[b]{0.5\linewidth}
  \centering
  \centerline{\includegraphics[width=\linewidth]{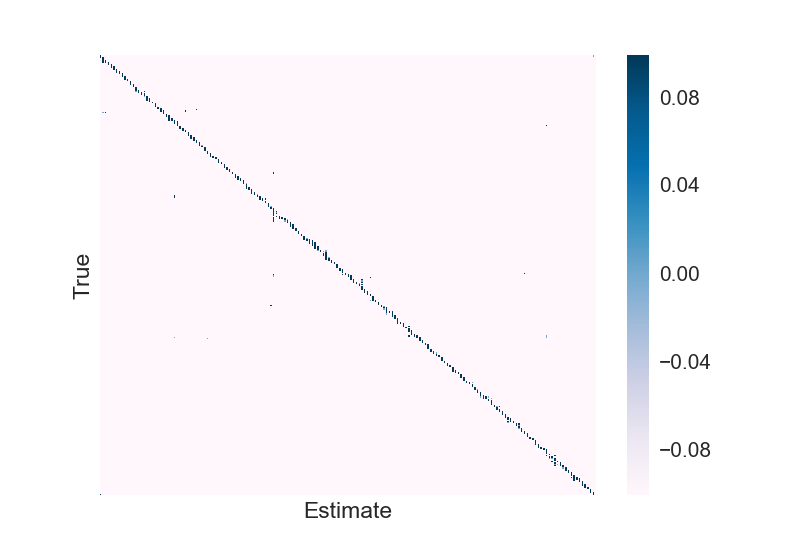}}
  \centerline{(a) One speaker.}\medskip
\end{minipage}%
\begin{minipage}[b]{0.5\linewidth}
  \centering
  \centerline{\includegraphics[width=\linewidth]{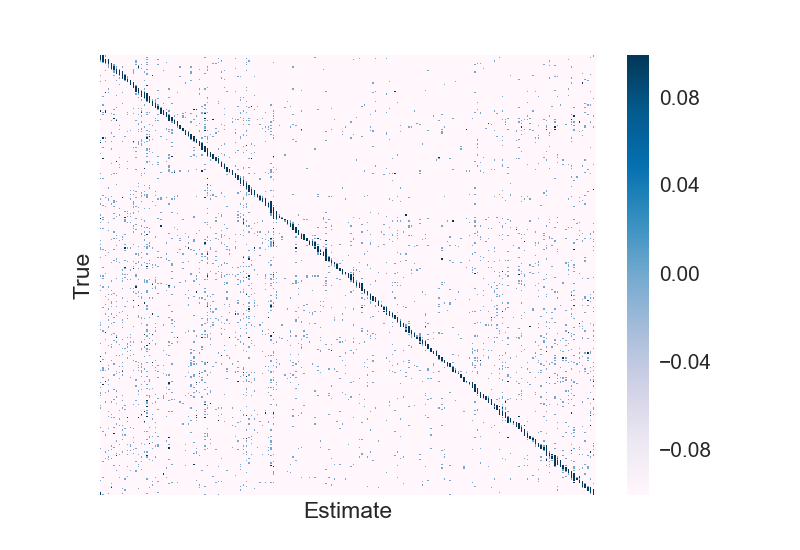}} 
  \centerline{(b) Two speakers.}\medskip
\end{minipage}
\vfill
\caption[]{Confusion matrices for localizing female speech with LEGO2 using a multiresolution approach. Improving the resolution decreases the number of outliers in the one-speaker case but not the two-speaker case.} 
\label{fig:confmatrixmutires}
\end{figure}

As before, multiresolution localization results in some improvement for one source but not for two sources (Table \ref{table:speechmulti}). We show the relevant confusion matrices in Figure \ref{fig:confmatrixmutires}: the lack of increase in performance can be explained by the fact that in the second round of localization the included directions are still strongly correlated and the only way to resolve the resulting ambiguities is through more constrained source models. Additionally, the set of correlated directions are not necessarily concentrated around the true direction which might explain the drop in accuracy for LEGO1. Overall, it seems the extra computation for the multiresolution approach does not bring about significant improvements compared to using a finer discretization.

\begin{table*}[t]
\caption{Error for speech localization with a multiresolution approach.}
\centering
\begin{tabular}{@{}lllllll@{}}
\toprule
&  \multicolumn{3}{c}{LEGO1}& \multicolumn{3}{c}{LEGO2}\\
& Accuracy & Mean & Per Source & Accuracy & Mean & Per Source\\\midrule
female speech & 96.94\% & $1.15^\circ$ & 96.94\%&  99.08\% & $0.70^\circ$ & 99.08\%\\
male speech &  86.00\% & $1.26^\circ$ & 86.00\%& 90.62\% & $0.95^\circ$ & 90.62\%\\
female/female & 17.88\%& $1.80^\circ$&56.66\% &  32.26\%& $1.08^\circ$ &  65.39\%    \\
female/male & 17.64\% & $1.87^\circ$ & 56.17\%&  29.06\%& $1.33^\circ$ & 63.47\%\\
male/male &  13.84\% &$2.19^\circ$ & 52.72\% & 20.22\%&  $1.64^\circ$ & 57.68\%\\
\bottomrule
\end{tabular}
\label{table:speechmulti}
\end{table*}

\begin{figure}[t]
\centering
\includegraphics[width=\linewidth]{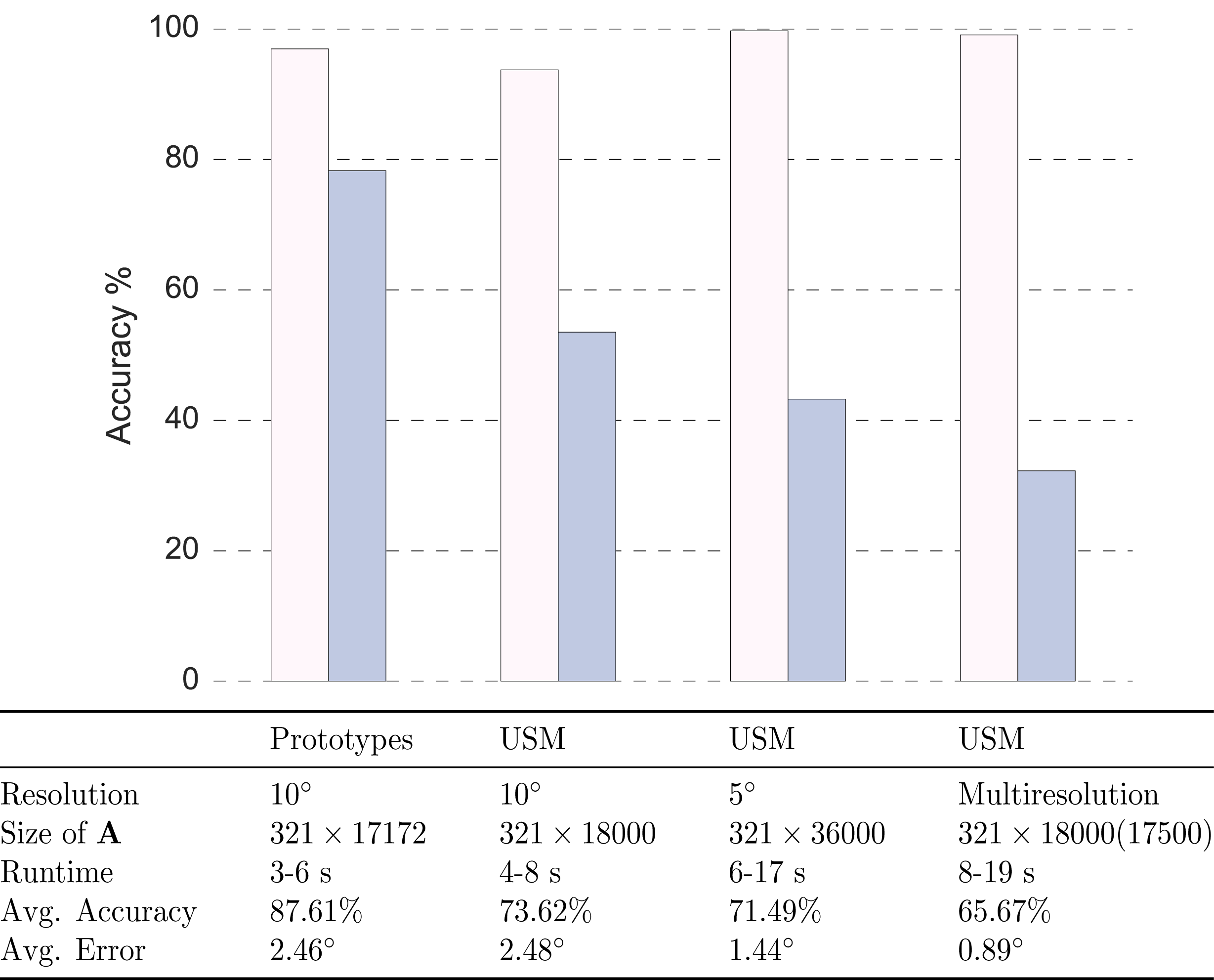}
\vspace*{-5mm}
\caption[]{Summary of localizing one (left) or two (right) female speakers using LEGO2.}
\label{fig:summary}
\end{figure}

Finally, in Figure \ref{fig:summary}, we show a summary of the performance of the different methods for localizing one or two female speakers using LEGO2 along with the average accuracy and error. Note that the results for prototypes use a smaller test set and that the error is lower bounded by the grid size. We also show the size of the model matrix $\mathbf{A}$ from \eqref{eq:mix} which contributes to the overall complexity of NMF as well as the actual runtime which depends on the machine. The figure suggests that overall using a USM and a $10^\circ$ resolution works well. For two-source localization, however, a good source model like prototypes is required.

\section{Conclusion}
Any scattering that causes spectral variations across directions enables monaural localization of one white source. On the other hand, more complex and interesting scattering patterns are needed to localize multiple sources. As shown by our ``random'' LEGO constructions, interesting scattering is not hard to come by. In order to localize general, non-white sources, one further requires a good source model. 

We demonstrated successful localization of one speaker using regularized NMF and a universal speech model. Both our LEGO scatterers were found to be superior in localization to a mannequin's HRTF. Finally, we stress that speech localization is challenging and note that the fundamental frequency of the human voice is below 300 Hz while the range of usable frequencies for our devices is above 3000 Hz. This discrepancy is responsible for outliers when localizing multiple speakers, a problem that can potentially be alleviated by increasing the size of the device or using sophisticated metamaterial-based designs. Perhaps a source model other than the universal dictionary could approach the performance of using prototypes.

Finally, we presented our results for anechoic conditions. Preliminary numerical experiments show that the current approach underperforms in a reverberant setting. This shortcoming is partly due to violations of our modeling assumptions. For example, in Eq. \eqref{eq:mixt}, the noise is assumed independent of the sources which is no longer true in the presence of reverberation. For practical scenarios it is thus necessary to extend the approach to handle reverberant conditions as well as to test the localization performance in 3D i.e., estimate both the azimuth and the elevation. For accurate localization in elevation, we expect that a taller device with more variation along the vertical axis would perform better. Since we only use one microphone, the number of ambiguous directions would likely grow considerably in 3D making the problem comparably harder. Other interesting open questions include blind learning of the directional transfer functions and understanding the benefits of scattering in the case of multiple sensors.  

\section{Acknowledgment}
We thank Robin Scheibler and Mihailo Kolund\v{z}ija for help with experiments and valuable comments. We also thank Martin Vetterli for numerous insights and discussions, and for suggesting Figure \ref{fig:responses}. This work was supported by the Swiss National Science Foundation grant number 20FP-1 151073, “Inverse Problems regularized by Sparsity”. 
\section{Disclaimer}
LEGO\rtm is a trademark of the LEGO Group which does not sponsor, authorize or endorse this work. 

\bibliographystyle{IEEEbib}

\end{document}